\documentclass[]{mn2e}
\usepackage{amssymb,psfig,graphicx}

\title[Star Cluster Formation in NGC 1569]{Star Cluster Formation and
Evolution in the Dwarf Starburst Galaxy NGC 1569}

\author[P. Anders et al.]{P. Anders$^1$\thanks{E-mail:
panders@uni-sw.gwdg.de}, R. de Grijs$^{2,3}$, U. Fritze--v.  Alvensleben$^1$, N. Bissantz$^4$\\
$^1$ Universit\"ats-Sternwarte, University of G\"ottingen, Geismarlandstr. 11, 37083 G\"ottingen, Germany, \\
$^2$ Institute of Astronomy, University of Cambridge, Madingley Road, Cambridge CB3 0HA \\
$^3$ Department of Physics \& Astronomy, The University of Sheffield, Hicks Building, Hounsfield Road, Sheffield S3 7RH \\
$^4$ Institut f\"ur Mathematische Stochastik, University of G\"ottingen, Lotzestr. 13, 37083 G\"ottingen, Germany
}

\date{Accepted ---. Received ---; in original form ---.}

\pubyear{2003}
\begin{document}
\maketitle

\begin{abstract}
We analyse multi-wavelength {\sl Hubble Space Telescope (HST)} observations of a large number of star clusters in
the nearby (post-) starburst dwarf galaxy NGC 1569. Their spectral energy
distributions (SEDs) cover at least the wavelength range from U to I in equivalent {\sl HST} filters, in most cases
supplemented by near-infrared data. Using our most up-to-date evolutionary
synthesis models of the G\"ottingen {\sc galev} code we determine ages,
metallicities, extinction values and masses for each individual cluster
robustly and independently. We confirm the youth of most of these objects. The
majority were formed in a very intense starburst starting around 25 Myr ago.
While there are two prominent ``super star clusters'' present in this galaxy,
with masses of $(5-16) \times 10^5 ~M_\odot$, almost all remaining clusters
are significantly less massive than an average Milky Way globular cluster, and
are generally consistent with open cluster-type objects. We determine the
cluster mass function from individual cluster masses derived by scaling the model SEDs of known mass to the observed cluster SEDs for each individual cluster. We find signs of a change in the cluster mass function as the
burst proceeds, which we attribute to the special conditions of star cluster
formation in this starburst dwarf galaxy environment.
\end{abstract}


\section{Introduction}
\label{intro.sec}

The dwarf starburst galaxy NGC 1569 (Arp 210, VII Zw 16, UGC 03056) has
attracted attention for almost 30 years, starting with the observations by
Hodge (1974) and de Vaucouleurs et al. (1974). Huge filamentary features are
seen in the outskirts of the galaxy, like the so-called ``H$\alpha$ arm'', as
well as bubbles and super-bubbles in all parts of the galaxy's main body (e.g.
Waller 1991, Heckman et al. 1995). This bubble structure has a complicated
velocity structure (Tomita et al. 1994) and is accompanied by signs of
galactic superwinds and outflows (Heckman et al. 1995, Della Ceca 1996),
caused by the massive energy input from collective supernova (SN) explosions
associated with the starburst. Whether the superwinds are strong enough to
remove a significant amount of material from the gravitational potential of
NGC 1569 is still being debated (this would predominantly remove the
high-metallicity SN ejecta, see e.g. Della Ceca 1996, Martin et al.
2002). Signs of recent star formation are seen along the bubble walls, which
is thought to be strong evidence for stochastic self-propagating star
formation (e.g., Gerola \& Seiden 1978, Seiden et al. 1984).

The properties of the two ``super star clusters'' (SSCs), usually called ``A''
and ``B'' (nomenclature from Arp \& Sandage 1985), have been studied in great
detail. First described by Arp \& Sandage (1985), a significant effort was
spent on characterising the properties of these clusters. Spectroscopic mass
estimates ($(2.3-8.3) \times 10^5 ~M_\odot$) were derived by Ho \& Filippenko
(1996) and Gilbert \& Graham (2001). Cluster ``A'' was resolved into a double
cluster with different stellar content in each of the components (de Marchi et
al. 1997, Buckalew et al. 2000 and Maoz et al. 2001, but see Gonz\'alez
Delgado et al. 1997 and Hunter et al. 2000 [``H00'']). To date the age
estimates of various groups agree fairly well, suggesting an age of cluster
``A'' of around 7 Myr (with probably a small age difference between the two
subclusters) and of around 10$-$20 Myr for cluster ``B'' (H00, Origlia et al.
2001, Maoz et al. 2001).

However, our knowledge of the remaining clusters is very limited. Only H00
have investigated a larger sample of star clusters in NGC 1569, but concentrate their
parameter studies on the SSCs. Hence only comparisons for photometric
performance can be made for the other clusters. Only the age of cluster ``no.
30'' (age $\approx$30 Myr) is presented elsewhere (Origlia et al. 2001,
nomenclature from H00).

This paper is part of an ongoing study, in which we will evaluate the impact
of the environment on the star cluster populations of galaxies with ongoing or
recent star cluster formation. While NGC 1569 is a gas-rich starburst dwarf
galaxy (Israel 1988), other environments, such as interacting galaxies of
various types and at various stages of interaction, will be studied with the
same methods in a homogeneous way.

\section{Observations and data preparation}
\label{obs.sec}

The data were retrieved from the {\sl Hubble Space Telescope (HST)} data archives, using the ESO/ST-ECF
``ASTROVIRTEL'' interface, and automatically calibrated using the standard
OPUS pipeline (On-the-fly Reprocessing), using the most up-to-date calibration
files available. A list of the retrieved WFPC2 and NICMOS data is provided in Table
\ref{obs.tab}. The original observations were taken in January 1996, October
1998 and February 1998 for PID 6111, 6423 and 7881, respectively.

\begin{table}
\caption[ ]{Overview of the observations of NGC 1569}
\label{obs.tab}
{\scriptsize
\begin{center}
\begin{tabular}{crccr}
\hline
\hline
\multicolumn{1}{c}{Filter} & \multicolumn{1}{c}{Exposure time} &
\multicolumn{1}{c}{Centre$^a$} & \multicolumn{1}{c}{PID$^b$} &
\multicolumn{1}{c}{ORIENT$^c$} \\
& \multicolumn{1}{c}{(sec)} & & & \multicolumn{1}{c}{($^\circ$)} \\
\hline
F336W & 2$\times$400    & PC   & 6423 & $-$86.172 \\
F380W & 2$\times$60     & PC   & 6111 &   170.066 \\
      &16$\times$600    & PC   & 6111 &   170.066 \\
F439W & 2$\times$40     & PC   & 6111 &   169.883 \\
      &16$\times$700    & PC   & 6111 &   169.883 \\
F555W & 2$\times$20     & PC   & 6111 &   169.714 \\
      &16$\times$500    & PC   & 6111 &   169.714 \\
      & 50, 2$\times$140& PC   & 6423 & $-$86.172 \\
      & 2$\times$300    & PC   & 6423 & $-$86.172 \\
F814W & 50, 2$\times$100& PC   & 6423 & $-$86.172 \\
      & 300             & PC   & 6423 & $-$86.172 \\
F110W & 10$\times$511.95 & NIC2 & 7881 & $-$139.691 \\
F160W & 10$\times$511.95 & NIC2 & 7881 & $-$139.691 \\
\hline
\end{tabular}
\end{center}
{\sc Notes:} $^a$ -- Location of the galactic centre; $^b$ -- {\sl HST}
programme identifier; $^c$ -- Orientation of the images (taken from the
image header), measured North through East with respect to the V3 axis
(i.e., the X=Y diagonal of the WF3 CCD $= + 180^\circ$).
}
\end{table}

Since the retrieved images are from three different proposals, the centerings
on the chip, and the orientations vary.

First, we divided the images into groups of the same passband/programme
combination. The images were checked for saturation effects, and the groups
were subdivided into subgroups with saturation and without saturation of the
brightest sources.

The images of each subgroup were combined, using the {\sc imalign} and {\sc
crreject/cosmicrays} tasks in {\sc iraf} \footnote{The Image Reduction and
Analysis Facility (IRAF) is distributed by the National Optical Astronomy
Observatories, which is operated by the Association of Universities for
Research in Astronomy, Inc., under cooperative agreement with the (U.S.) National
Science Foundation. {\sc stsdas}, the Space Telescope Science Data Analysis
System, contains tasks complementary to the existing {\sc iraf} tasks. We used
Version 2.3 (June 2001) for the data reduction performed in this paper.}. The
subgroups were then rotated, their pixel sizes matched, aligned and trimmed
using the appropriate {\sc iraf} routines.

The final fields-of-view (FoVs) are: 577 $\times$ 577 PC pixels (26\farcs54
$\times$ 26\farcs54) for the small FoV (which includes NICMOS coverage), and
777 $\times$ 787 PC pixels (35\farcs74 $\times$ 36\farcs20) for the larger FoV
common only to the WFPC2 observations. These FoVs correspond to areas of 283
$\times$ 283 pc and 381 $\times$ 386 pc, respectively, at an adopted distance
for NGC 1569 of 2.2 Mpc (see e.g. Israel 1988)

\subsection{Source selection}

\begin{figure}
\includegraphics[angle=-90,width=\columnwidth]{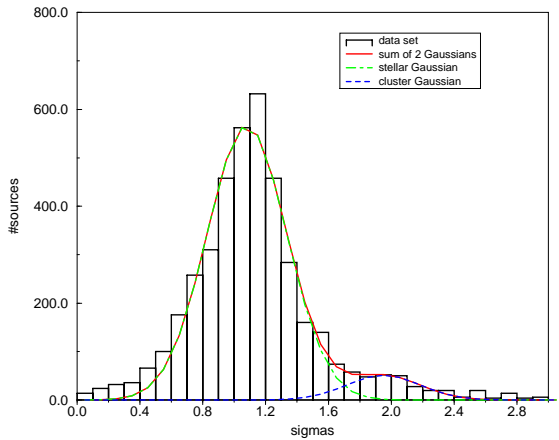}
\caption{Distribution of the Gaussian $\sigma$'s of the detected sources. Two
Gaussians were fitted to this distribution, representing stars and cluster
candidates, respectively.}
\label{fig_sig}
\end{figure}

\begin{figure*}
\includegraphics[width=\linewidth]{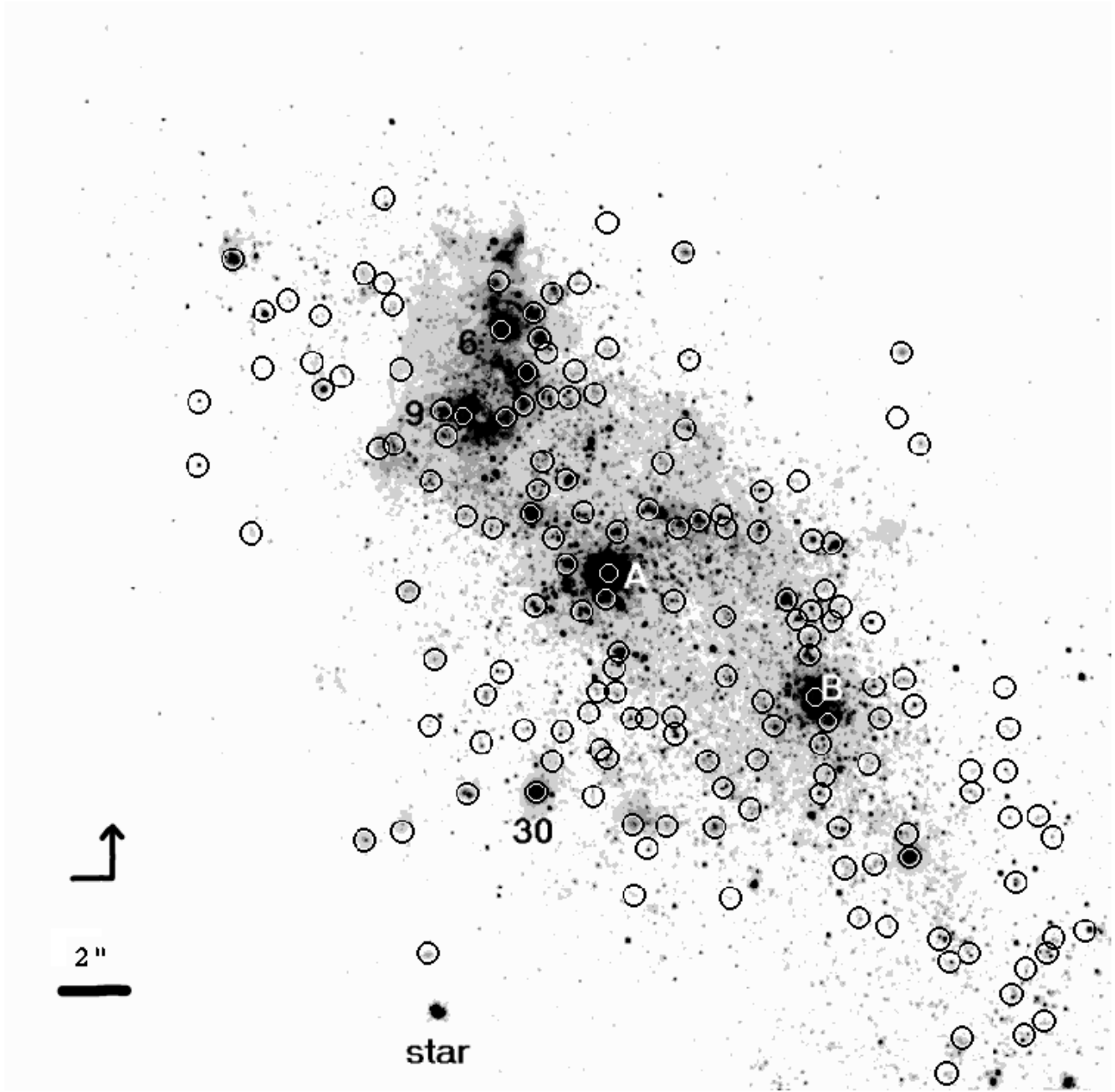}
\caption{Image of the inner part of NGC 1569 (our small FoV) with the
positions of the clusters marked. Some clusters (and one apparent star) are
labelled. North is marked by the arrow, east is indicated by the line perpendicular to the arrow.}
\label{fig_image}
\end{figure*}

Objects were selected using a version of {\sc daofind} (Stetson 1987), running
under {\sc idl}. Subsequently, the source sizes were estimated by fitting
Gaussian profiles to the sources' intensity distributions. ``Point-like sources'' may contain bright
stars in NGC 1569 as well as foreground stars in the Galaxy. At least one
obvious foreground star is visible in the lower right-hand corner of Fig.
\ref{fig_image}. Sources with $\sigma \le$ 1.7 pixels (equivalent to 0.83 pc
at the distance of NGC 1569) were rejected as being point-like sources. This
limit was chosen after an analysis of the distribution of Gaussian $\sigma$'s,
shown in Fig. \ref{fig_sig}. A sum of two Gaussians was fitted to this
distribution, one representing point-like sources and the other cluster-like
(hence significantly extended) sources. The rejection limit at $\sigma$ = 1.7
pixels was chosen conservatively to get a clean cluster sample. As pointed out
by Whitmore et al. (1999) a broadening of the (intrinsic) point-spread function (PSF)
might occur at deep exposures due to jitter and breathing effects. Additional
broadening might be caused by crowding of star clusters, although this
seems to be a minor issue in NGC 1569 where the star clusters are usually well
separated (see Fig. \ref{fig_image}), and during the alignment and
combination of large numbers of single images. These effects are most likely
responsible for the pronounced peak around $\sigma \simeq$ 1.2 pixels, hence
$\approx$0.4 pixel larger than the corresponding value for a pure WFPC2 PSF.
This procedure also removes blends of bright stars and clusters from our
sample, for which accurate cluster photometry would be impossible.

We believe that this method gives reliable results, since we are only interested in
relative size estimates and we measure the sizes of the clusters in a homogeneous way.

An initial list of cluster candidates was created by cross-correlating the source
lists from ``saturated'' F555W and F814W (comparable to V-band and I-band)
images. Subsequently, additional bright sources from the other passbands were
added.

Finally, all candidate clusters were examined visually. Spurious detections,
remaining cosmic rays covering more than a single pixel and obvious remaining single
bright stars or blends of single stars and clusters were rejected in
this step.

The number of star clusters resulting from this procedure was 168 in the
small FoV, and 179 in the larger FoV.

\subsection{Photometric calibration}
\label{photcal.sec}

The coordinates from the source lists obtained in the previous section were
used as the centres for aperture photometry, based on {\sc daophot} routines
adapted for {\sc idl}, in all passbands. Standard apertures of 5 pixel source
aperture and 5$-$8 pixel sky annuli were adopted, corresponding to 2.45 pc and
2.45$-$3.9 pc at the distance to NGC 1569 of 2.2 Mpc. Visual inspection allowed us to adjust these standard
apertures where necessary (for almost half of the clusters, the
apertures and sky annuli needed to be increased), to include a maximum
fraction of the source flux and to avoid unrelated features in both the source
aperture and the sky annulus. After performing the aperture photometry,
magnitudes from saturated sources were replaced by the correct values from the
unsaturated images.

The full data tables containing the integrated photometry and analysis results of all clusters are available, at http://alpha.uni-sw.gwdg.de/$\sim$panders/data/NGC1569/.

Note that, in order to determine model magnitudes from our model spectra (see Sect. \ref{algorithm.sec}), we adopt the full filter response functions, {\it including the red leak in F336W}. Therefore,
we need not attempt to correct for the red leak of the F336W filter. In general, the
impact of the red leak is negligible for spectra dominated by early-type stars
such as a starburst spectrum (e.g. Eskridge et al. 2003, de
Grijs et al. 2003a). In addition, any ambiguity originating from the red leak
is resolved by our spectral energy distribution (SED) analysis tool, which takes into account the whole SED.

\begin{figure}
\includegraphics[angle=-90,width=\columnwidth]{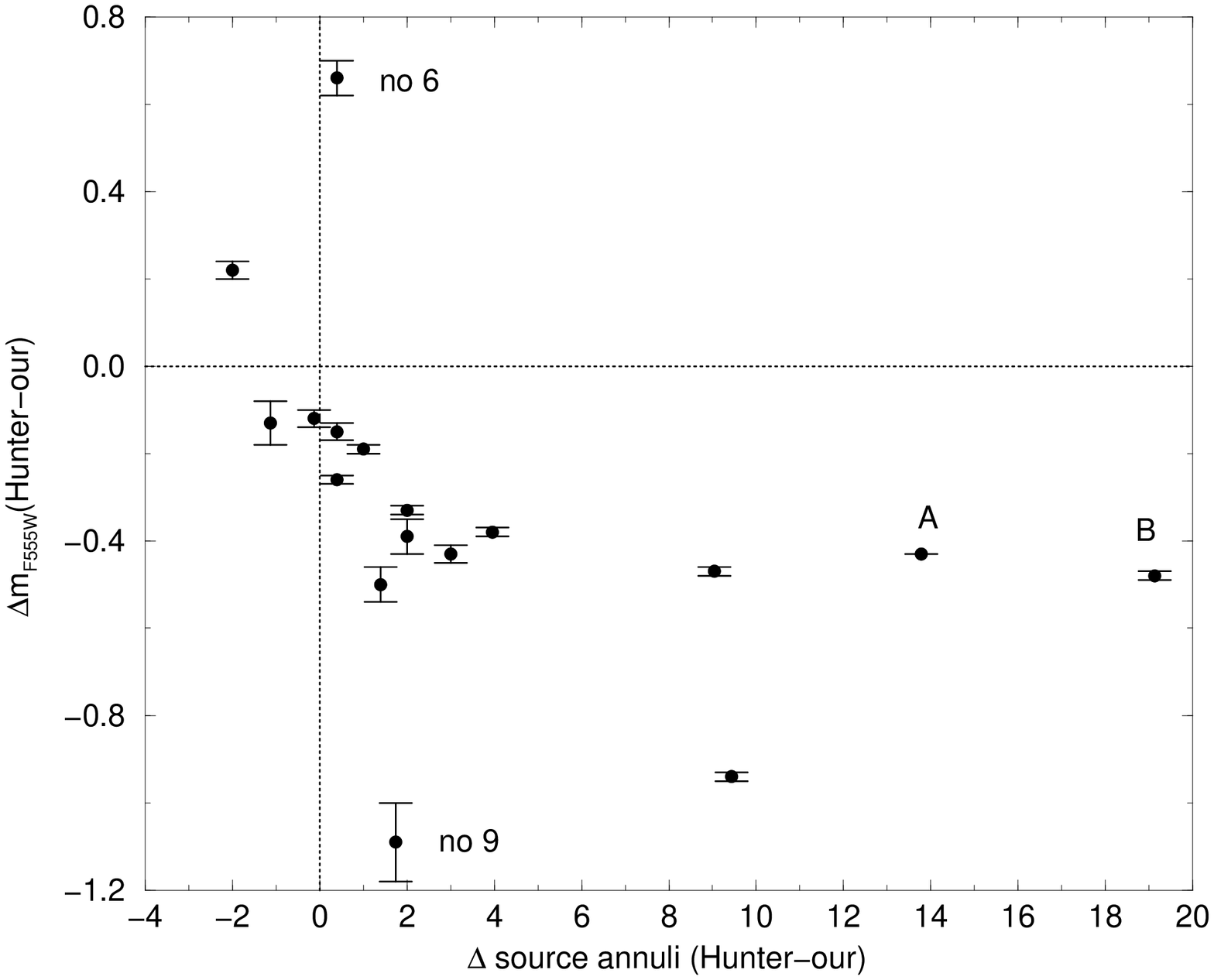}
\caption{Comparison of the cluster photometry obtained from H00 and in this
work. Displayed is the dependence of the magnitude difference on the
difference of the source apertures used. Naming of 4 individual clusters is
following H00. Lines are included to guide the eye.}
\label{fig_Hunter}
\end{figure}

From the 47 clusters of H00, only 17 are matched in our source list.
Roughly a dozen of the H00 clusters lie outside of our final FoVs from the combined {\sl HST} data, and hence are
not included in this work. Another 9 H00 clusters are in the close vicinities of larger
clusters, and are therefore rejected by our selection criteria; the sizes
of most of these clusters, as estimated by H00, are close to our lower size limit
of 1.7 pixel. In addition, the source selection criteria are not described in
detail by H00, and hence comparison is difficult or impossible. By comparing
the photometry of the clusters in common, after correction for the different distance
moduli and extinction values considered, we find the H00 values to be
brighter, in general, than ours, which can be attributed to different
apertures used for the photometry (mostly larger in the case of H00), see Fig.
\ref{fig_Hunter}. H00 attempted to extract the contribution of
nebular emission from their science images, in particular of H$\alpha$ emission, using appropriate narrow-band filters to estimate the approximate H$\alpha$ line intensity and its continuum level.
However, further details are not given in their paper. This makes comparison
difficult, but since we account for nebular emission in our models the
parameter analysis is not affected. The error bars in Fig. \ref{fig_Hunter}
are calculated from the photometric errors in H00 and from our work. In
general, the median photometric errors for our clusters are of the order of
0.05 mag except for the F336W filter where they are around 0.1 mag. The
photometric errors considered in this paper include Poissonian noise in the
background and in the source flux itself, and variations in the background
level.

Only in a few cases the correlation of brighter clusters with larger source
annuli in the H00 sample does not hold. The SSCs are surrounded by smaller
substructures and superimposed on generally high background levels, where
reliable aperture photometry is difficult, in terms of both source and background
fluxes. In such cases, the results are very sensitive to the exact positions
and sizes of the source and background annuli.

We will now discuss the few true ``outliers'' in Fig. \ref{fig_Hunter} in more detail. 
H00 cluster ``no. 9'' lies close to another bright source (probably a bright
star), hence contamination of the source flux of the cluster by the other
source is likely.

Cluster ``no. 6'' is located in a region with very strong gas emission
(and is a strong emitter itself because of its young age; we determine an age of 4
Myr), according to the false-colour image including H$\alpha$ emission of H00 (their Fig. 2).
This gas emission might lead to misassignment of flux to either the source or
the background flux.

These differences in the integrated cluster photometry between the H00 values
and ours reflect the advantages of a visual inspection of each cluster in the complicated environments associated with ongoing starbursts,
compared to automated cluster photometry. Non-cluster features, such as crowding
or extended gas emission, are taken into account more reliably.

In Anders et al. (2003) we show that SEDs of
young clusters exhibit a ``hook'' near the {\sl B} band (see also Fig. \ref{fig_SED}). Tracing this hook is of
the highest importance for the determination of accurate cluster parameters. The deep observations in F380W and the additional F336W observations are vital for
tracing this hook. In addition, having near-infrared (NIR) data available further improves the
analysis. For the majority of our clusters we have NIR data at hand. Hence,
the observations available to us for NGC 1569 are ideally suited to produce accurate
cluster parameters.

\subsection{Sample completeness}

\begin{figure}
\includegraphics[angle=-90,width=\columnwidth]{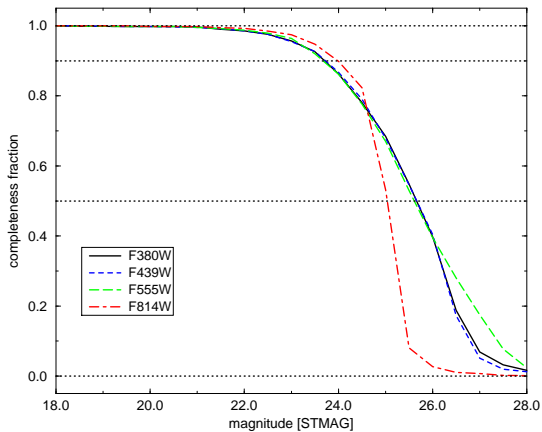}
\caption{Completeness curves for NGC 1569. The different line styles refer to different passbands, as indicated in the legend. These completeness curves apply to the small FoV. The horizontal lines indicate 0, 50, 90 and 100 per cent completeness. Magnitudes are given in STMAG, the standard {\sl HST} zero-point system, based on a flat reference spectrum.}
\label{fig_comp}
\end{figure}

The completeness of the exposures was determined by adding artificial sources
of known brightness to the saturated science images, and subsequent source
identification. This analysis was performed using the appropriate {\sc iraf}
and {\sc idl} tasks in a similar way as for the source selection of the real
clusters. By cross-correlating the list of retrieved objects with the input artificial
sources, the fraction of recovered artificial sources was determined. The
results for the small FoV are shown in Fig. \ref{fig_comp}. The completeness
limits for the large FoV are similar.

These completeness curves were corrected for the effects of blending or
superposition of multiple randomly placed artificial PSFs as well as for the
superposition of artificial PSFs on top of genuine objects. For the analysis
in this paper, completeness limits at 90 per cent will be considered.

\section{Parameters of the young clusters}

\subsection{Cluster analysis algorithm}
\label{algorithm.sec}

We applied our maximum-likelihood algorithm to the magnitudes of the cluster
candidates found in Section \ref{photcal.sec}. This algorithm and several
tests of it are described in Anders et al. (2003). Only a summary of the
method is given here.

We use the evolutionary synthesis models of our {\sc galev} code, described in
detail in Schulz et al. (2002), with important additions regarding the
treatment of gaseous emission in the early stages of cluster evolution
presented in Anders \& Fritze -- v. Alvensleben (2003). We also calculate
model magnitudes with internal dust extinction, by adopting the starburst
extinction law of Calzetti et al. (2000), assuming a foreground screen
geometry. Our extinction estimates are therefore, strictly speaking, lower limits. Galactic extinction is taken into account by dereddening the
observations using the appropriate Galactic extinction values from Schlegel et al. (1998).

Our models are based on stellar isochrones from the Padova group which include
the thermally-pulsing AGB-phase shown to be vital to correctly predict the
colours of clusters with ages between 200 Myr and 1 Gyr (see Schulz et al.
2002). Throughout the paper we adopt a Salpeter initial mass function (IMF)
with lower mass limit $M_{\rm low}$ = 0.15 $M_\odot$ and upper mass limit $M_{\rm up}
\approx$ 50-70 $M_\odot$, determined by the upper mass
limit of the Padova isochrones ($M_{\rm up}$ = 50 $M_\odot$ for the highest
metallicity Z=0.05, $M_{\rm up} \approx$ 70 $M_\odot$ for the lower
metallicities). Adopting a different IMF, however, affects the derived (absolute) masses of the clusters, but the
effect on the other parameters, and on the {\it relative} mass distribution, is negligible. The
mass offset by assuming a different IMF can easily be derived analytically.
The spectral library used is given in Lejeune et al. (1997, 1998). The
emission line coefficients for low metallicities are taken from Izotov et al.
(1994, 1997, 1998), and from Stasi\'nska (1984) for metallicities $\ge$ 0.008.
The Lyman-continuum photon output was calculated by Schaerer \& de Koter
(1997), and recently confirmed by Smith et al. (2002), see Anders \& Fritze - v. Alvensleben (2003) for details.

Our models assume a well-populated IMF, which is an over-simplified assumption for
systems with small numbers of bright stars, as shown e.g. by Cervi\~no et al.
(2002) and Cervi\~no \& Valls-Gabaud (2003). Small-number statistics and stochastic effects for bright
stars, such as Wolf-Rayet stars or supergiants, introduce additional model
magnitude dispersions, which scale inversely with the mass of the cluster. However, no
complete study for all magnitudes and input parameters has yet been performed.

We construct SEDs from these models with an
age resolution of 4 Myr for ages from 4 Myr up to 2.36 Gyr, and with 20 Myr
resolution for older ages (up to a maximum age of 14 Gyr). The extinction
resolution is $\Delta$E(B$-$V) = 0.05 mag, for E(B$-$V) in the range of
0.0$-$1.0 mag. The adopted metallicities are [Fe/H]=$-$1.7, $-$0.7, $-$0.4, 0.0, 0.4, as
given by the Padova isochrones used (for a general description of the
stellar models see Bertelli et al. 1994 and Girardi et al. 2000; for details
about the isochrones in our models see Schulz et al. 2002).

When comparing our model SEDs with the observed SEDs we first determine the
mass of the cluster by shifting the model SED onto the observed SED. This
shift is equivalent to scaling the model's mass to the cluster mass.

Each of the models is now assigned a certain probability to be the most
appropriate one, determined by a likelihood estimator of the form $p \sim
\exp(-\chi^2)$, where $
\chi^2=\sum{\frac{(m_{\rm obs}-m_{\rm model})^2}{\sigma^2_{\rm obs}}}$. Clusters with
unusually large ``best'' $\chi^2$ are rejected, since this is an indication of
calibration errors, features not included in the models (such as Wolf-Rayet
star dominated spectra, objects younger than 4 Myr etc.) or problems due to
the limited parameter resolutions. The lower cut-off is set to a total probability
= $10^{-20}$, corresponding to a ${\rm \chi^2_{best} \ge 46}$. The total
probability per cluster is then normalised.

Subsequently, the model with the highest probability is chosen as the
``best-fitting model''. Models with decreasing probabilities are summed up
until reaching 68.26 per cent total probability (= 1 $\sigma$ confidence
interval) to estimate the uncertainties of the best-fitting model. These
uncertainties are in fact upper limits, since their determination does not
take into account effects like the existence of several ``solution islands'' for
one cluster (such as e.g. the age-metallicity degeneracy, see Section \ref{agemet.sec}), and
discretisation in parameter space.

Several passband combinations (containing at least 4 passbands) were used for
the analysis, to minimise the impact of statistical effects on the errors and
calibration errors. A minimum of 4 passbands is required to determine the 4
free parameters age, metallicity, extinction and mass independently (see also
Anders et al. 2003, de Grijs et al. 2003a). We caution that these passband
combinations must not be biased to contain mainly short-wavelength filters or
mainly long-wavelength filters. Coverage of the entire optical wavelength
range, if possible with the addition of ultra-violet (UV) {\it and} 
NIR data, is most preferable (de Grijs et al. 2003a). We select the passband
combinations starting with all available filters, and then rejecting passbands
starting with the shallowest exposures and exposures not covering the entire
combined FoV.

Only clusters with observational errors $\le$ 0.2 mag in all passbands of a
particular combination were included to minimise the uncertainties in the
results. For each combination, the best-fitting models and the associated
parameter uncertainties were determined. For a certain cluster all best-fitting
models (and the associated uncertainties), originating from the different
passband combinations, were compared. For each of these best-fitting models the
product of the relative uncertainties (${\rm \frac{age^+}{age^-} \times
\frac{mass^+}{mass^-} \times \frac{Z^+}{Z^-}}$) was calculated (the superscripts
indicate the upper limits ($^+$) and the lower limits ($^-$), respectively)
The relative uncertainty of the extinction was not taken into account, since
the lower extinction limit is often zero. For each cluster, the data set with the lowest value
of this product was adopted as the most representative set of parameters (and
parameter uncertainties). In cases where the
analysis converged to a single model, a generic uncertainty of 30 per cent was assumed for
all parameters in linear space, corresponding to an uncertainty of
$^{+0.1}_{-0.15}$ dex in logarithmic parameter space. See also de Grijs et al.
(2003a,b) for an application of this algorithm to NGC 3310 and NGC
6745, and Anders et al. (2003) for a theoretical analysis of its reliability.

The figures presented in this paper are based on the overall best values.

\subsection{Parameter distributions}

In total, we identified and analysed 144 clusters in the small FoV, and 157
clusters in the large FoV. All of these clusters meet the minimum error
criterion in at least one passband combination, while 24/22 additional
clusters (small FoV/large FoV) were rejected by this criterion. Every accepted
cluster was matched with representative single stellar population models by our algorithm.

We mainly used 3 passband combinations for the analysis: all 7 passbands
(``{\sl 7mag}''), the combination of F380W, F439W, F555W, F814W, and F160W
(``{\sl UBVIH}'') and the most restricted combination of F380W, F439W, F555W,
and F814W (``{\sl UBVI}''). In the small FoV we managed to match 62 clusters
using {\sl 7mag}; 24 additional clusters were matched using {\sl UBVIH} (since
either their F336W or their F110W magnitudes had errors larger than 0.2 mag).
The remaining 58 clusters in the final sample for the small FoV could only be
analysed without NIR information. These latter clusters are
mostly concentrated towards the corners of this FoV, where due to the rotation
applied to the NICMOS images there is no NIR information available. To
summarise, if NIR information is available, passband combinations including
the NIR filter NICMOS F160W give the best results in almost all cases,
based on their relative uncertainty product.

For the small FoV we checked the uncertainties inherent to the analysis
routine. We find median uncertainties of 1 step in metallicity (as a reminder, the
metallicities used are [Fe/H]=$-$1.7, $-$0.7, $-$0.4, 0.0, 0.4), 0.1 mag in
extinction E(B-V), a factor of 3 in age (corresponding to a logarithmic
uncertainty of 0.5), and a mass uncertainty of a factor of 2.3 (corresponding
to a logarithmic uncertainty of 0.35). This is in good agreement with
theoretical tests of our algorithm (see Anders et al.
2003 for more details). We have chosen the bin sizes in the following
figures based on these uncertainty estimates.

\subsubsection{Parameter distributions of the entire sample in the small FoV}
\label{small_FoV.sect}

\begin{figure}
\includegraphics[angle=-90,width=\columnwidth]{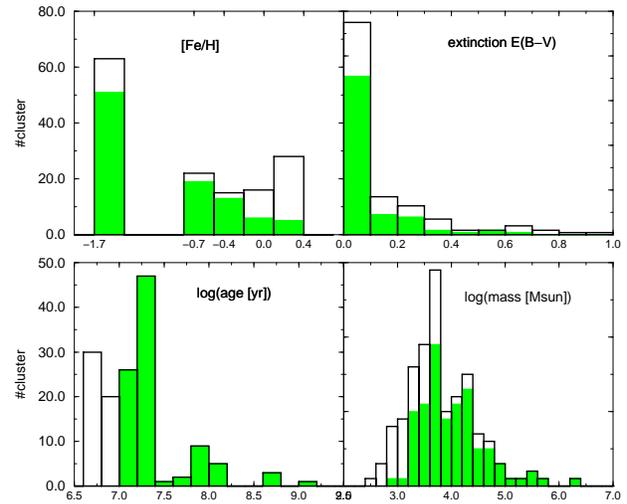}
\caption{Parameter distributions of the clusters in the small FoV; open histograms: all
ages, shaded histograms: only ages $>$ 8 Myr.}
\label{fig_par_young_old}
\end{figure}

In Fig. \ref{fig_par_young_old} the derived parameter distributions of the
clusters in the small FoV are shown (open histograms).

\begin{figure}
\includegraphics[angle=-90,width=\columnwidth]{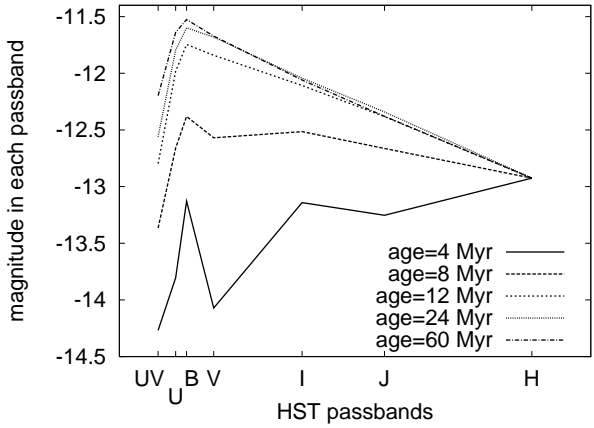}
\caption{Model SEDs for Z=0.004 and 5 different young ages. SEDs are shifted
to coincide at the {\sl H} band.}
\label{fig_SED}
\end{figure}

The metallicity distribution (Fig. \ref{fig_par_young_old}, [Fe/H]) is
dominated by significantly subsolar-abundance clusters. The high-metallicity
part is dominated by the youngest clusters (ages $\le$ 8 Myr), as can be seen
from a comparison with the shaded histograms. We attribute this to the
age-metallicity degeneracy. To further quantify the associated effects we
analysed the fractions of clusters without NIR data (and hence less reliable
parameters, especially the metallicity is fixed most effectively using NIR
data) in certain age and metallicity bins. We found the youngest ages (at 4
Myr $\approx$63 per cent, at 8 Myr $\approx$50 per cent) and the highest metallicities
(for Z=0.02=Z$_\odot$ $\approx$56 per cent, for Z=0.05 $\approx$68 per cent) to be
dominated by clusters for which no NIR photometry with uncertainties $\le$ 0.2 mag was
available. In the other bins, the average fraction of clusters without NIR data is
around 30 per cent.

Almost all clusters were produced in the last 25 Myr in a very intense burst.
Only 21 clusters ($\approx$15 per cent) are older. This is consistent with the
existing evidence regarding the ages of the SSCs, and the dynamical age of
certain morphological features like arcs and superbubbles (de Vaucouleurs et
al. 1974, Waller 1991, Tomita et al. 1994, Heckman et al. 1995). While these
morphological features are thought to be evidence for self-propagating
star-formation, our age determination does not provide further proof of this scenario, since no
spatial concentration of clusters at any given age is observed.

In Figure \ref{fig_SED} we present a number of model SEDs for the
low-metallicity environment of NGC 1569 and the young ages typical for its
star clusters. With typical observational errors of 0.05-0.1 mag these SEDs
are clearly distinguishable. The UV range is crucial for this distinction, in
agreement with our theoretical analysis in Anders et al. (2003). The
deep observations of NGC 1569 in the F380W filter (approximately equivalent to the {\sl U} band) are
vital for the accuracy of our results.

Colour-magnitude diagram analyses yield more extended starburst histories for
NGC 1569 starting up to 150 Myr ago, and ending around 5$-$10 Myr ago, with
ongoing low-level star formation for the last 1.5 Gyr (Vallenari et al. 1996, Greggio
et al. 1998, Aloisi et al. 2001). Since such extended burst scenarios are not
supported by our determinations of the cluster ages, we rather associate this
with the low-level star cluster formation for clusters with log(age) $\ge$ 7.4
and the secondary peak of star cluster formation at around 100 Myr ago.

The extinction towards the NGC 1569 clusters is low: 73 per cent of the clusters have E(B$-$V) $\le$ 0.1.
Only 10 clusters ($\approx$7 per cent) have E(B$-$V) $\ge$ 0.5, and all of these are young
(ages $\le$ 16 Myr, with most of them as young as 4 Myr).

The masses of the cluster candidates are low compared to Galactic globular
clusters, which have a Gaussian shaped mass distribution with $\langle
log(M_{\rm GC} [{\rm M}_\odot]) \rangle_{\rm MW} \simeq 5.5$ and $\sigma (log(M_{\rm GC}
[{\rm M}_\odot])_{\rm MW}) \simeq 0.5$ (Ashman et al. 1995). In the case of the star
cluster sample in NGC 1569 we are likely observing a system of open
cluster-type objects rather than globular cluster progenitors. Only 4 objects
have masses in excess of log(mass [$M_\odot$])=5.47.

\begin{figure}
\includegraphics[angle=-90,width=\columnwidth]{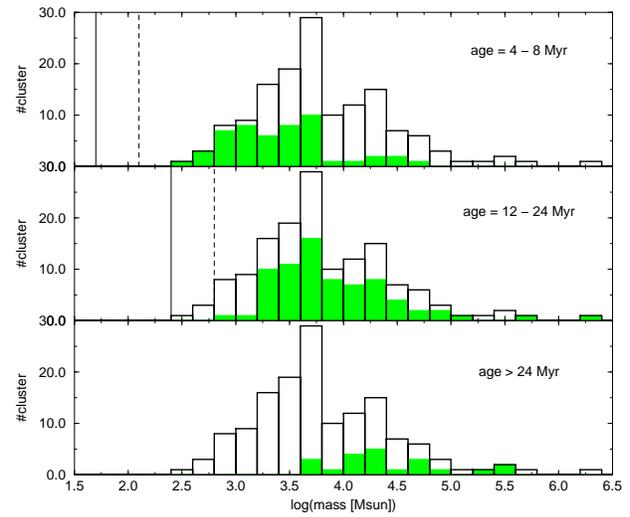}
\vspace*{+0.2cm}
\caption{Mass functions of NGC 1569 cluster candidates in three age bins
(shaded histograms; ages as indicated in each panel) and the total mass function (open
histograms). Vertical lines indicate completeness limits (see text for
details).}
\label{fig_MF_age}
\end{figure}

An even more remarkable result is shown in Fig. \ref{fig_MF_age}. There is
significant evidence that the clusters formed at the onset of the burst are, on
average, more massive than the clusters formed more recently. The vertical
solid lines in the top two panels indicate the completeness limits for ages of
4 Myr and 24 Myr, respectively. The dashed lines indicate completeness if one
assumes an additional drop in completeness of a yet another 1.0 mag due to the
visual examination, because of a possible bias to reject preferentially
fainter clusters. We would require an additional drop of 1 mag to explain the
decrease at log(mass) $<$ 3.2 in the second age bin's shaded histogram.
Alternatively, this drop might be caused by disruption of the lowest-mass clusters on
time-scales as short as 25 Myr.

\begin{figure}
\includegraphics[angle=-90,width=\columnwidth]{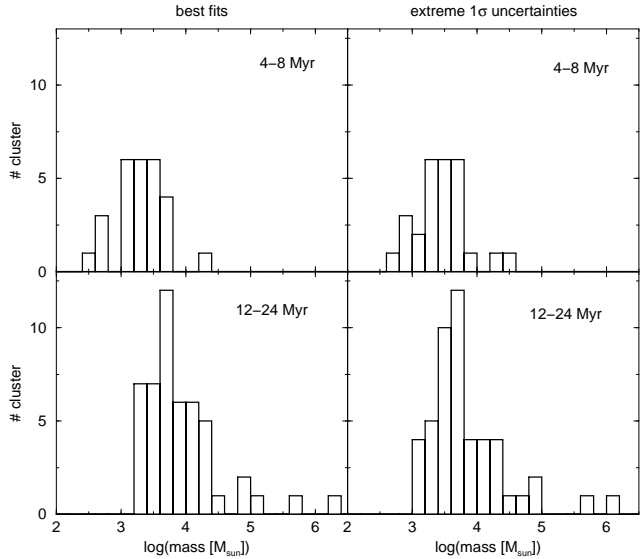}
\vspace*{+0.2cm}
\caption{Mass functions of NGC 1569 cluster candidates in two age bins (as
indicated), with age uncertainties entirely in this age bin. Left panels: best
values. Right panels: Extreme cases allowed by uncertainty estimates (young
ages: upper limits; older ages: lower limits). See text for details.}
\label{fig_MF_agesig}
\end{figure}

To investigate the significance of this change in mass function we
display, in Fig. \ref{fig_MF_agesig}, a subset of clusters with age
estimates that are entirely within the respective age bins. The right-hand panels
show the most extreme configuration allowed by the 1$\sigma$ uncertainties of
our mass estimates, where we assume that all clusters in the younger age bin
have masses at the upper limit of the 1$\sigma$ mass uncertainty, and all
clusters in the older age bin have masses at the lower limit of the 1$\sigma$
mass uncertainty range. An excess of clusters with log(mass) $\ge$ 3.8 is still
clearly visible. Quantitatively, this excess is significant at roughly the 10
$\sigma$ level and cannot be explained even by worst-case Poisson-noise scenarios.

While the behaviour at the low-mass end remains debatable due to
incompleteness effects, the excess of clusters with log(mass) $\ge$ 3.8 in the
intermediate age bin is significant. We will discuss a number of possible
explanations for this excess in Section \ref{formation.sect}.

\subsubsection{The ``well-known'' clusters}

Since most information in the literature is available for the clusters ``A'',
``B'' and ``no. 30'', we compare our determinations
for these clusters to those found in the literature in Table 2. The masses are
taken from Ho \& Filippenko (1996) (``HF96'') and Gilbert \& Graham (2001)
(``GG01''). The ages are from H00 and Origlia et al. (2001).

The ages of the two SSCs determined in this work agree very well with those
from the literature. The agreement is not as good
for cluster ``no. 30'' (our determination indicates a significantly older age than the
literature value), but since no uncertainties are given in the literature, a
direct comparison is difficult. However, since cluster ``no. 30'' is located
at some distance from the bar of the galaxy, where the vast majority of star
clusters is concentrated, a formation earlier than the major burst is not
unlikely.

We determine higher masses than inferred from kinematic studies. Three
reasons are possible:
\begin{enumerate}

\item We systematically determine ages that are too old, and due to the rapid
changes in the mass-to-light (M/L) ratio at these early stages the resulting
masses are too high. This scenario is not supported by our age uncertainties.

\item We adopted an incorrect stellar IMF. If the low-mass slope of the IMF is
shallower than Salpeter (see e.g. Kroupa et al. 1993), our mass estimates
need to be reduced by a factor of roughly 2 (see de Grijs et al. 2003b).

\item The masses calculated from the measured velocity dispersions are underestimated.

\end{enumerate}

The kinematic masses of GG01 are about a factor of 2 smaller than ours, which
is most likely within the uncertainties inherent to both methods (e.g. including the
uncertainty in the IMF, evolutionary synthesis
uncertainties, parameter uncertainties originating from the observation-model
comparison, and uncertainties in the kinematic masses). A mass of ${\rm 3.3 \times 10^5
~M_\odot}$ was derived by HF96 for SSC A. However, they did not take into
account the substructure of this cluster, nor its impact on the velocity
dispersion. They also assumed a different distance modulus, and a fixed sigma to correct
for the impact of red supergiants, rather than measuring it from the
autocorrelation function, as done by GG01 (although HF96 claim to give a lower mass
limit only). Nevertheless, both studies assume complete virialisation of the
clusters, which is likely not the case for clusters as massive and as young as
these two SSCs (hence the dynamical mass estimate is likely an underestimate).
A cluster's half-mass relaxation time is given by $T [{\rm yr}] = 8 \times 10^5
\frac{n^{1/2} \cdot R^{3/2}}{\langle m \rangle^{1/2} \cdot (\ln(n)-1)}$, with $n$ the
number of stars in the cluster, $R$ its radius in pc and $\langle m \rangle$ the median mass of a
star in the cluster. The radii are taken from de Marchi et al. (1997): 1.6 pc
for SSC A1, 1.8 pc for SSC A2/B. We assume that the half-light radius approximates the half-mass radius for young clusters (cf. de Grijs et al. 2002a), a median mass of a star in the cluster is
$\approx$0.3 M$_\odot$ (for a well-populated IMF ranging from 0.15 ${\rm
M_\odot}$ to 70 ${\rm M_\odot}$ [upper mass limit given by Padova isochrones
for low metallicity]), masses are adopted from GG01 ($3.9 \times 10^5 {\rm M}_\odot$
for A1, $4.4 \times 10^5 {\rm M}_\odot$ for A2 and $2.3 \times 10^5 {\rm M}_\odot$ for B),
numbers of stars in the clusters are calculated from the total cluster mass
and the characteristic stellar mass. These values result in half-mass
relaxation times of 250-400 Myr, well in excess of the expected ages of these
clusters by more than a factor of 20. However, this relaxation model
does not account for effects of mass segregation (for observational evidence
of significant {\sl ab initio} mass segregation in a sample of young LMC clusters see de Grijs
et al. 2002b), or radial dependences of the relaxation time-scales. Hence
uncertainties are large. Our data do not allow to discriminate between these
sources of uncertainties.

The best-fitting metallicity for NGC 1569 derived from CMD analyses is ${\rm
[Fe/H] = -0.7}$ (Greggio 1998, Aloisi 2001, both using Padova tracks). This
agrees well with spectroscopic abundance measurements by Kobulnicky \&
Skillman (1997) and Devost et al. (1997). Both teams measure abundances of around 12
+ [O/H] = 8.2, corresponding to [Fe/H] = $-$0.7. We find both SSC B and no. 30 
best matched by models with [Fe/H] = $-$0.4, and hence comparable to the literature
values cited within the uncertainties associated with the methods. SSC A is
best matched by a model with the lowest metallicity available ([Fe/H] =
$-$1.7). This might reflect the uncertain character of this star cluster,
which may consist of two subclusters (de Marchi 1997), but certainly contains
two very distinct populations (Gonz\'alez Delgado et al. 1997, de Marchi et
al. 1997, H00, Maoz 2001): Wolf-Rayet stars and red supergiants. Hence a
simple single stellar population model is probably not appropriate for this cluster. In
addition, Wolf-Rayet stars are not specifically marked in the Padova
isochrones, and the treatment of red supergiants by the Padova group differs
from that of e.g. the Geneva group. The treatment of these
stars is not yet beyond debate.

\begin{table}
\caption[ ]{Comparison of parameters of clusters with literature values}
\label{ssc.tab}
{\scriptsize
\begin{center}
\begin{tabular}{lcccc}
\hline
\hline
\multicolumn{1}{c}{cluster} & \multicolumn{1}{c}{Average} & \multicolumn{1}{c}{Uncertainties} & \multicolumn{1}{c}{Best values} & \multicolumn{1}{c}{Uncertainties} \\
& \multicolumn{1}{c}{Literature} & \multicolumn{1}{c}{Literature} & \multicolumn{1}{c}{This work} & \multicolumn{1}{c}{This work}\\
\hline
age [Myr]\\
\hline
SSC A$^a$ & 7 & ~4$-$20 & 12 & ~8$-$16 \\
SSC B & 15 & 10$-$30 & 12 & 12$-$28 \\
No. 30 & 30 & $-$ & 92 & ~28$-$112 \\
\hline
\hline
\multicolumn{1}{c}{cluster} & \multicolumn{1}{c}{HF96} & \multicolumn{1}{c}{GG01} & \multicolumn{1}{c}{Best values} & \multicolumn{1}{c}{Uncertainties} \\
& \multicolumn{1}{c}{} & \multicolumn{1}{c}{} & \multicolumn{1}{c}{This work} & \multicolumn{1}{c}{This work}\\
\hline
mass $[M_\odot]$\\
\hline
SSC A$^a$ & $3.3 \times 10^5$ & $8.3 \times 10^5 ~^b$ & $1.6 \times 10^6$ & $(1.1-2.1) \times 10^6$ \\
SSC B & $-$ & $2.3 \times 10^5$~~ & $5.6 \times 10^5$ & $(5.6-8.8) \times 10^5$\\
No. 30 & $-$ & $-$ & $3.6 \times 10^5$ & $(2.8-6) \times 10^5$\\
\hline
\end{tabular}
\end{center}
{\sc Notes:} $^a$ Analysis in this work converged to 1 model only; 30 per
cent uncertainties are assumed. $^b$ Sum of both subcomponents.
}
\end{table}

\subsubsection{Comparison with other subsamples and the age-metallicity
degeneracy}
\label{agemet.sec}

\begin{figure}
\includegraphics[angle=-90,width=\columnwidth]{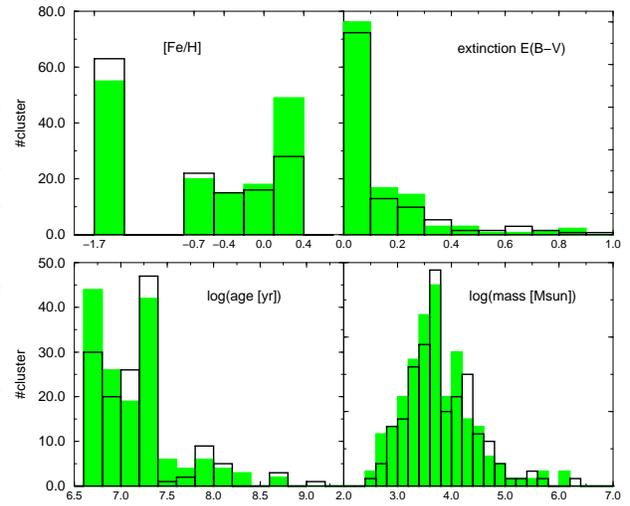}
\caption{Parameter distributions of the two FoVs: open histograms = small FoV (including
NICMOS coverage), shaded histograms = large FoV (without NICMOS coverage).}
\label{fig_par_both}
\end{figure}

Figure \ref{fig_par_both} compares the parameter distributions of clusters in
the small FoV with the ones from the large FoV. The major difference between
these two FoVs is not just the spatial coverage, but -- more importantly -- for all
clusters in the large FoV the NIR information is either unavailable or omitted.

While the extinction and log(mass) distributions are fairly similar, the
metallicity and age distributions show a significant shift. The clusters in
the large FoV seem to have, on average, higher metallicity and younger ages.
While these parameters are expected to correlate, a closer inspection of the
data reveals a different situation. Comparing the same clusters in both the
small and the large FoVs shows that of the clusters classified as having
super-solar metallicity in the large FoV only 48 per cent are classified as
solar/super-solar in the small FoV. Instead, some 33 per cent of them were
originally classified as having the lowest possible metallicity [Fe/H] =
$-$1.7. In addition, of the 28 clusters in the small FoV's super-solar
metallicity bin, 19 (68 per cent) have no NIR information available (due to
the rotated NICMOS FoV). Single stellar population models run close (and
partly intersect each other) in colour evolution for different metallicities
at early phases. As widespread super-solar abundances seem fairly implausible
in this kind of dwarf galaxy, we strongly suspect these results to be a clear
sign of the age-metallicity degeneracy. This degeneracy can only be broken by
using NIR observations, in addition to UV-optical data, in particular to constrain
the metallicity of the clusters properly. From this comparison we strongly
recommend the use of NIR facilities for multi-band photometry to determine
reliable cluster parameters from broad-band photometry, at least for clusters
as young as in our sample.

\subsubsection{Investigating certain metallicity range restrictions}

\begin{figure}
\includegraphics[angle=-90,width=\columnwidth]{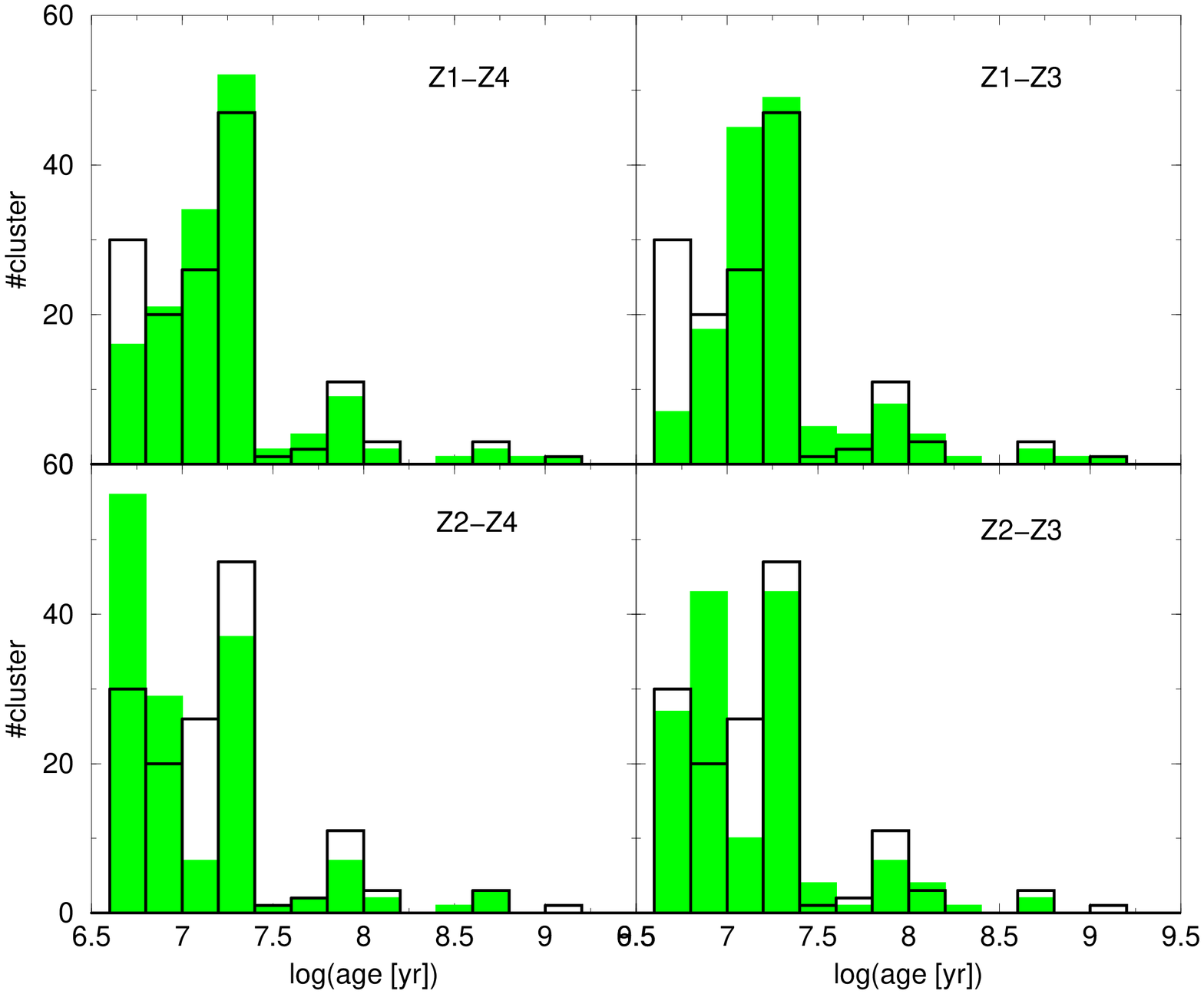}
\includegraphics[angle=-90,width=\columnwidth]{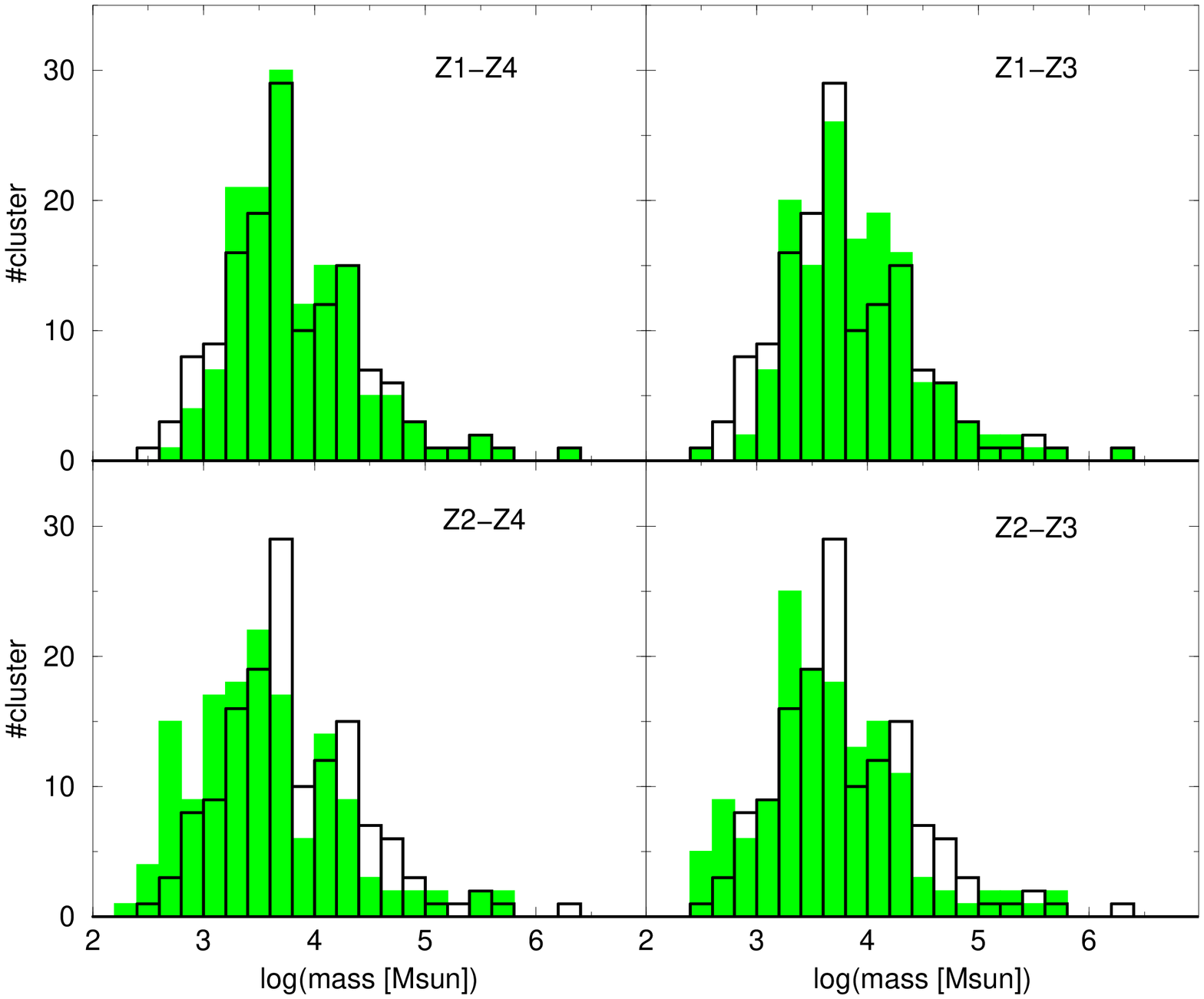}
\caption{Comparison of parameters for certain metallicity restrictions
(shaded histograms) for the small FoV. Upper panel: log(age) distributions, lower panel:
log(mass) distributions. The allowed metallicity ranges are shown in the
panels (Z1 $\equiv$ [Fe/H]=$-$1.7, Z2 $\equiv$ [Fe/H]=$-$0.7, Z3 $\equiv$
[Fe/H]=$-$0.4, Z4 $\equiv$ [Fe/H]=0, Z5 $\equiv$ [Fe/H]=+0.4). The results for
no metallicity restriction are shown as open histograms, for comparison.}
\label{fig_Zres}
\end{figure}

To evaluate the robustness of our determinations, we performed cluster
analysis with restricted metallicity ranges. The results for
the log(age) and log(mass) distributions are shown in Fig. \ref{fig_Zres}.
While the rejection of the super-solar/solar metallicity range is
justified physically (since large numbers of high-abundance clusters are not
expected in a dwarf galaxy environment, such as in NGC 1569), the rejection of
the lowest metallicity is a more theoretical exercise (since subsolar
metallicity ranges cannot be omitted {\sl a priori}).

These tests confirm the general properties obtained without restrictions in
metallicity space, proving the robustness of our analysis method. The extinction (not shown) and
log(mass) distributions are very similar, with only minor changes. The
log(age) distributions confirm the onset of the major burst 24 Myr ago, and a
minor burst around 100 Myr ago. The detailed structure of the major burst,
however, depends on the metallicities allowed. Rejection of high
metallicities leads to a depopulation of the youngest age bins. This is
related to the study of the impact of the availability of NIR data on the
derived parameters (Section \ref{small_FoV.sect}). Rejection of the
lowest metallicity leads to a distribution skewed to younger ages, to balance
the mean higher metallicity.

\begin{figure}
\includegraphics[angle=-90,width=\columnwidth]{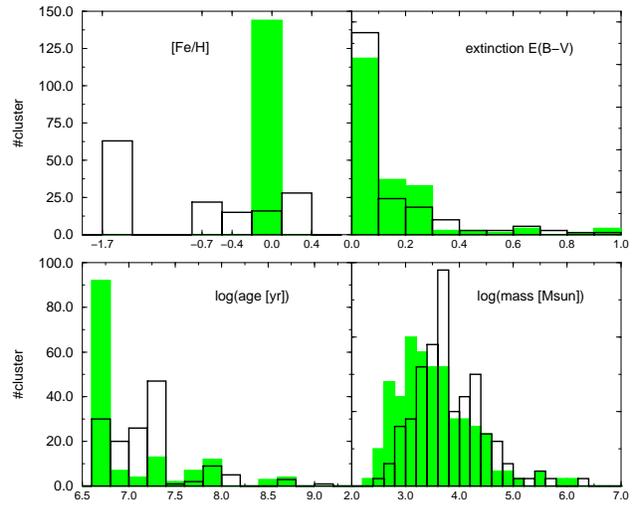}
\caption{Comparison of parameters, derived with all metallicities allowed
(open histograms) and with metallicity fixed to solar (shaded histograms)}
\label{fig_comp_Z4}
\end{figure}

The importance of robust, independent metallicity determinations can also be
seen from Fig. \ref{fig_comp_Z4}, which shows the parameter distributions for
the clusters in the small FoV, assuming solar metallicity
(shaded histograms). The open histograms are the distributions from
the same sample without such restrictions, for comparison. Major changes are
visible in the log(age) (a clear shift towards younger ages for the
metallicity-restricted values) and log(mass) distributions (towards lower
masses, which is a consequence of the younger ages, in conjunction with the
associated rapidly changing M/L ratios).

To summarise, we have analysed a large sample of star clusters in NGC 1569 by
comparing the observed cluster SEDs with an extensive grid of model SEDs to
determine the cluster ages, metallicities, internal extinction values and
masses in a robust and homogeneous way. We have presented the best-fitting values for the clusters, and
compared them to the results with restricted parameter spaces or wavelength
coverage. We conclude that we can determine ages (and hence the star cluster
formation history; bursty, with a major peak starting 25 Myr ago, and a minor
peak around 100 Myr ago), masses (similar to open cluster-type objects),
extinction values (the average internal extinction is found to be low) and
metallicities (significantly sub-solar; however, the impact of the
age-metallicity degeneracy is clearly seen, especially for clusters without
NIR data) robustly, with well-understood uncertainties.

\section{Cluster disruption vs. fading}
\label{sect.dis_fad}

\begin{figure}
\includegraphics[angle=-90,width=\columnwidth]{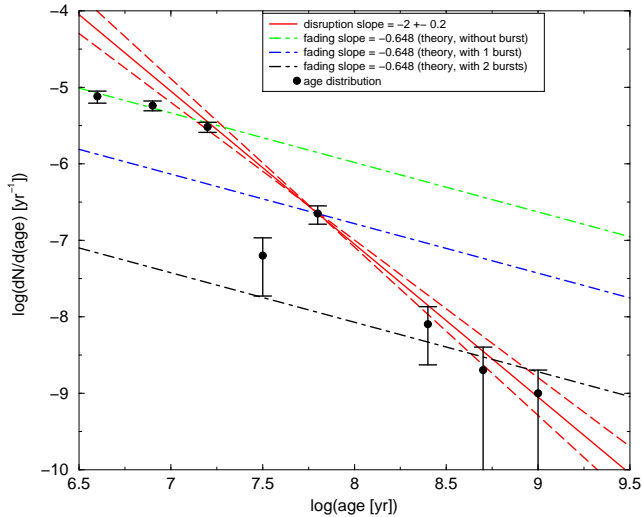}
\caption{Cluster age distribution, affected by fading and cluster disruption.
Linear relations are indicated, and slopes given in the legend; see text for details.}
\label{fig_disr_age}
\end{figure}

\begin{figure}
\includegraphics[angle=-90,width=\columnwidth]{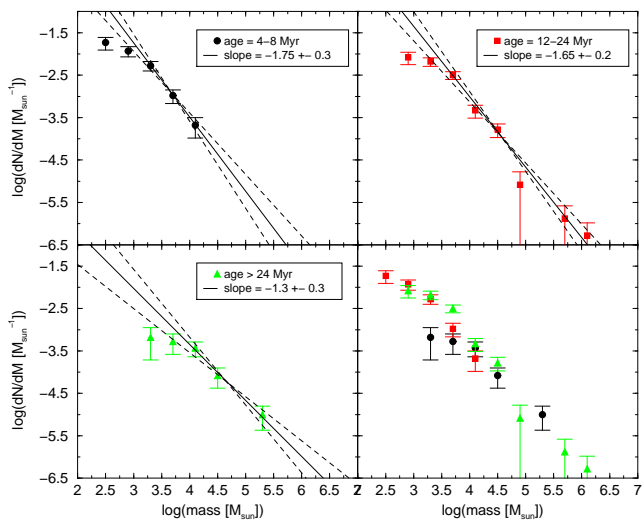}
\caption{Cluster mass distribution for 3 different age bins, as indicated in
the legends. Best-fitting slopes (and uncertainties) shown, as described in
Section \ref{sect.dis_fad}.}
\label{fig_disr_mass}
\end{figure}

We applied the method of Boutloukos \& Lamers (2003) regarding
cluster detectability limited by fading due to stellar evolution and cluster
disruption to our cluster sample from the small FoV. The results are shown in
Figs. \ref{fig_disr_age} and \ref{fig_disr_mass}. Since the total number
of clusters is small, statistics are relatively poor, but consistent.

Figure \ref{fig_disr_age} shows the number of clusters formed per year, as a
function of log(age). The fading lines are based on the slope given by
Boutloukos \& Lamers (2003) for the V-band (slope $\alpha_{\rm fade}$ = $-$0.648) and are shifted vertically to best
match the data. The disruption line is a fit to
the data points with log(age) $\ge$ 7.2 (with slope $\alpha_{\rm disr}$ = $-$2 $\pm$ 0.2).

Since the Boutloukos \& Lamers (2003) method is based on the
assumption of a constant cluster formation rate, which is not valid in our
case, the interpretation of this figure is ambiguous:

\begin{enumerate}

\item As a fading part (short-dashed line, for log(age) $\le$ 7.2) and a
disruption part (solid line), without any burst (which is unrealistic,
compared to Fig. \ref{fig_par_young_old}).

\item With enhanced cluster formation in the recent past, which is only
affected by fading (upper dot-dashed line) and offset from the low-level
cluster formation fading line (middle dot-dashed line, for log(age) $\ge$
7.5), with subsequent cluster disruption (solid line).

\item With fading only, but with 2 bursts (upper dot-dashed line [log(age)
$\le$ 7.2] and middle dot-dashed line [log(age) $\simeq$ 7.8]) and low-level
cluster formation (lower dot-dashed line).

\end{enumerate}

We treat the burst as a temporarily constant cluster formation rate
(Boutloukos \& Lamers 2003; de Grijs, Bastian \& Lamers 2003), shifting the theoretical fading
lines to fit our data in the respective age intervals.

Figure \ref{fig_disr_mass} shows the number of clusters formed per
$M_\odot$ for 3 age bins. Due to small-number statistics the slope determinations
are fairly uncertain, but comparable. The average slope, however, is shallower
than expected (we find $\alpha_M \simeq$ $-$1.6, other studies find slopes
around $-$2, see de Grijs et al. 2003b for a comparative compilation), but
this might again be caused by the small-number statistics. Since the slopes
for all age bins are comparable, we conclude that cluster disruption cannot
yet have played a significant role, not even for the oldest clusters, which seems
plausible in a low-density environment such as in NGC 1569. The lowest masses
in each bin are clearly affected by incompleteness effects. The errors included in Fig. \ref{fig_disr_mass}
are derived from Poissonian statistics in the respective mass bins.

\section{The physical requirements for star cluster formation}
\label{formation.sect}

In Section \ref{small_FoV.sect} we observed a change in the mass function
towards lower masses as the burst of star cluster formation proceeded. We
performed a number of tests, to check whether this result (as shown in Fig.
\ref{fig_MF_age}) is physically true or whether it might be only due to
small-number statistics. First we counted the clusters in the mass bins
log(mass)=3.2$-$3.8 and log(mass)$>$3.8, subdivided into the age bins
age=4$-$8 Myr and age=12$-$24 Myr (as a reminder, we only have an age
resolution of 4 Myr). We find 24 clusters in the younger+less massive bin, 7
clusters in the younger+more massive bin, 37 clusters in the older+less
massive bin, and 34 clusters in the older+more massive bin. If we assume the
younger bins to be correct, we can estimate from the older+more massive bin
what we expect for the older+less massive bin: we would expect $\approx$ 117
clusters to be in the older+less massive bin, which is roughly 13 $\sigma$ away from
the observed value, if one assumes purely Poissonian statistics. Secondly, we
performed a KS-test to estimate the probability that the mass functions in
both age bins are drawn from the same distribution. We find a probability of
only 10 per cent if we adopt a lower mass cut-off at log(mass)=3.2 (to account
for completeness effects), and a probability of only 14 per cent if we adopt a
lower mass cut-off at log(mass)=3.5 (further reducing the completeness limit),
hence both distributions are significantly different. Both tests
show the significance of our results. There are several possible reasons for
this change in the mass function with time.

One reason might be a relation between cluster masses and the available gas
reservoir and the possible exhaustion of the gas reservoir by the first
generation of star clusters in the burst, leaving only a small amount of gas
available for the star clusters formed more recently. While this seems
plausible, there {\sl is} still a large amount of gas available. Israel (1988)
estimated the amount of gas left in the galaxy to be mass(H{\sc i}) = ${\rm
1.1 \times 10^8 ~M_\odot}$ and mass(H$_2$) = ${\rm 2 \times 10^7 ~M_\odot}$,
while the total mass of the galaxy is estimated to be mass(total) = ${\rm 3.3
\times 10^8 ~M_\odot}$. Hence NGC 1569 is not gas-poor. It might still be a
matter of the distribution of gas, and of the gas density. Taylor et al.
(1999) studied the distribution of CO (and hence H$_2$) in the centre of NGC
1569. They concluded that while there is a sufficient amount of H$_2$
(mass(H$_2$) $\simeq$ ${\rm 7.7 \times 10^6 ~M_\odot}$), this mass is
distributed over an area $\approx$200 pc across, and hence not dense enough to
form new clusters. The observed giant molecular clouds (GMCs), on the other
hand, are not massive enough to form new SSCs assuming any reasonable overall
star formation efficiency. It is noteworthy that the observed GMCs are not in
the vicinity of the SSCs, but close to observed H{\sc ii} regions, and hence near
the most actively star-forming regions. The H{\sc i} maps of Stil \& Israel
(2002) also indicate a depression of (neutral) gas near the SSCs (see also Greve et al. 2002), and a clumpy
higher-density ridge (with intensity peaks East and West of the SSCs,
$\approx$ 15\arcsec away from the SSCs, and thus just outside our FoV) along the
galaxy's major axis, together with extended diffuse emission. The total amount
of H{\sc i} gas is estimated to be $\approx$ ${\rm 1.3 \times 10^8 ~M_\odot}$,
but spread across a large area.

As suggested by the observations of Stil \& Israel (1998), the starburst in
NGC 1569 might be triggered by the passage of a nearby H{\sc i} cloud
(projected distance $\simeq$ 5 kpc) with mass = ${\rm 7 \times 10^6
~M_\odot}$. There seems to be an H{\sc i} ridge connecting NGC 1569 to this
companion (Stil \& Israel 1998, their fig. 3), supporting a scenario in which the H{\sc i} cloud passed close to
NGC 1569 in the recent past, on time-scales equivalent to the burst duration,
i.e. of the order of few tens of Myr. The ram pressure compression of the
interstellar medium (ISM) during the approach of the companion, and the
absence of this compression during the time after perigalacticum, is another
possible explanation for the change in the mass functions with time. This
explanation is supported by observational (e.g. de Grijs et al. 2001, 2003b)
and theoretical evidence (Ashman \& Zepf 2001, Elmegreen 2002) of enhanced
average star cluster masses and star formation efficiencies caused by
interaction-induced ram pressure.

A final possible, and in our opinion most likely, origin might be the strong
radiation field caused by the large number of newly formed massive stars in
the beginning of the burst and the follow-up energy input into the ISM by
SNe. Waller (1991) and Origlia et al. (1998) estimate the number of
SNe during the burst to be of order 2000$-$25,000 SNe/Myr. This might
not only cause an unusually high dust temperature (34 K; e.g. Hunter
et al. 1989, Lisenfeld et al. 2002) and the powering of the galaxy's strong thermal
X-ray halo and bipolar outflow (Heckman et al. 1995, Della Ceca et
al. 1996, Martin et al. 2002), but it might also prevent the assembly of
larger molecular clouds due to heating of the ISM and pressure by UV photons
and SN ejecta, since the collapse time-scale of a Jeans-instable cloud
increases with its mass. Observed GMCs are likely to have collapsed while
shock fronts, caused by the outflow of material, passed through high-density
warm material (shock-cooling, see Taylor et al. 1999), and therefore driven by conditions
not available in all regions of the galaxy.

\section{Summary}

We interpret multi-colour {\sl HST} data of star clusters in the dwarf (post-)starburst
galaxy NGC 1569 with dedicated evolutionary synthesis models and a robust
analysis method to determine ages, metallicities, internal extinction values
and masses of individual clusters independently. We conclude that we are
observing a mainly young, recently formed, perhaps partially still forming,
star cluster system, which mainly consists of objects considerably less
massive than average globular clusters in the Milky Way. The extinction within
NGC 1569 towards the clusters is found to be low, the metallicity distribution
wide-spread, which is thought to be at least partially due to the
age-metallicity degeneracy. These results are consistent with previous
studies, but enlarge the sample of star clusters analysed in NGC 1569 by a factor of 4.

We confirm the bursty character of star cluster formation in NGC 1569, with a
major burst starting some 25 Myr ago, possibly triggered by a passing H{\sc i}
cloud, and approximately continuous, low-level star cluster formation at earlier times.
We detect a significant lack of high-mass clusters formed at the end of the
burst, compared to clusters formed earlier in the burst. While the reason
for this is still unclear, we consider three possibilities (or any interplay
of them):

\begin{enumerate}

\item the absence of a sufficient amount of gas, or gas density, to form new
(massive) star clusters;

\item the ram pressure caused by the companion H{\sc i} cloud is apparent predominantly
during the companion's approach, and not during its recession;

\item the heating of the ISM due to a strong radiation field and SN
ejecta originating from the clusters formed in the beginning of the burst,
preventing the assembly of massive molecular clouds by photo-ionisation and
turbulence.

\end{enumerate}

We believe all three scenarios to be of relevance, but attribute the highest
importance to the latter one.

From a technical point of view we conclude that:

\begin{enumerate}

\item The commonly used procedure to assume a generic metallicity (and
extinction) for all clusters is dangerous, since this affects
the resulting age (and mass) distributions significantly.

\item With only optical passbands available the age-metallicity degeneracy
largely precludes the determination of reliable ages and masses. While it {\it
is} partially possible to correct for this, it introduces additional
uncertainties. A more reliable way is by using additional NIR observations.

\end{enumerate}

In the near future we will apply our methods to different star-forming
environments, such as interacting galaxies of various types and in various
stages of interaction. This will improve our understanding of the impact of
the environment on the formation, evolution and disruption processes of
recently formed star clusters, and their relation with old star clusters, like
the well-studied globular clusters in the Milky Way.

\section{Acknowledgments} This paper is based on archival observations with
the NASA/ESA {\sl Hubble Space Telescope}, obtained at the Space Telescope
Science Institute, which is operated by the Association of Universities for
Research in Astronomy (AURA), Inc., under NASA contract NAS 5-26555. This
paper is also partially based on ASTROVIRTEL research support, a project
funded by the European Commission under 5FP Contract HPRI-CT-1999-00081. This
research has made use of NASA's Astrophysics Data System Abstract Service.
This research has also made use of the NASA/IPAC Extragalactic Database (NED)
which is operated by the Jet Propulsion Laboratory, California Institute of
Technology, under contract with the National Aeronautics and Space
Administration. PA is partially funded by DFG grant Fr 911/11-1. PA also
acknowledges partial funding from the Marie Curie Fellowship programme
EARASTARGAL ``The Evolution of Stars and Galaxies'', funded by the European
Commission under 5FP contract HPMT-CT-2000-00132. We would like to thank Henny
Lamers for many fruitful discussions and the hospitality of the University of
Utrecht, as well as the anonymous referee for many useful suggestions.

\end{document}